\documentclass[prl, reprint]{revtex4-1}

\usepackage{amsmath, amssymb}
\usepackage{times, txfonts}
\usepackage[utf8]{inputenc}
\usepackage[english]{babel}
\usepackage{graphicx}
\usepackage{hyperref, lipsum}

\renewcommand{\Im}{\mathop{\mathrm{Im}}}
\renewcommand{\Re}{\mathop{\mathrm{Re}}}

\begin{document}

\date{December 18, 2017}

\title{Polariton chimeras: Bose-Einstein condensates with intrinsic chaoticity\\
  and spontaneous long-range ordering}

\author{S.~S.~Gavrilov}

\affiliation{Institute of Solid State Physics, RAS, Chernogolovka,
  142432, Russia,}

\affiliation{A.~M.~Prokhorov General Physics Institute, RAS, Moscow,
  119991, Russia,}

\affiliation{National Research University Higher School of Economics,
  Moscow 101000, Russia}

\begin{abstract}
  The system of cavity polaritons driven by a plane electromagnetic
  wave is found to undergo the spontaneous breaking of spatial
  symmetry, which results in a lifted phase locking with respect to
  the driving field and, consequently, in the possibility of internal
  ordering.  In particular, periodic spin and intensity patterns arise
  in polariton wires; they exhibit strong long-range order and can
  serve as media for signal transmission.  Such patterns have the
  properties of dynamical chimeras: they are formed spontaneously in
  perfectly homogeneous media and can be partially chaotic.  The
  reported new mechanism of chimera formation requires neither
  time-delayed feedback loops nor non-local interactions.
\end{abstract}

\maketitle

\textit{Introduction.}---Dynamical \emph{chimeras} represent a novel
concept in nonlinear science.  In the case of continuous media they
can be defined as long-range patterns that (i) arise spontaneously in
perfectly homogeneous environment and (ii) comprise regular and
chaotic subsystems~\cite{Panaggio15}.  A chimera state may collapse
into a fully ordered state or, conversely, undergo turbulent
destruction.  In a sense, ``order'' and ``chaos'' act as two balanced
sides of a single essence.  Discovered by Kuramoto in the field of
oscillator networks~\cite{Kuramoto02,Abrams04,Acebron05}, chimera
states have recently been evidenced in various systems in nonlinear
optics~\cite{Larger15,Larger13}, mechanics~\cite{Martens13},
chemistry~\cite{Tinsley12}, and neurophysiology~\cite{Andrzejak2016}.
Here we show that chimeras can arise in systems of locally interacting
Bose particles and involve strong long-range ordering of such systems.

We consider a cavity-polariton system driven by a plane
electromagnetic wave. Cavity polaritons are short-lived composite
bosons formed owing to the strong coupling of excitons (electron-hole
pairs in semiconductors) and cavity photons; they are excited
optically and emit light~\cite{Weisbuch92,Yamamoto-book-2000}.  Under
coherent pumping, their macroscopic states are treated as highly
nonequilibrium Bose condensates (\cite{Elesin73,Haug83}) obeying a
nonlinear Schr\"{o}dinger equation~\cite{Kavokin-book-07}.  Today,
growing attention is paid to pattern formation due to spin-sensitive
interaction of polaritons.  In particular, the circular-polarization
degree of the light wave transmitted through or emitted by the
microcavity can be varied in space and time~\cite{Sarkar10, Adrados10,
  Kammann12, Anton15, Cilibrizzi16, Solnyshkov09-j}.  Spin patterns
usually form as a result of artificial or random structural disorder
or space-dependent driving field~(\cite{Shelykh08-j, Shelykh08-prl,
  Gavrilov12-prb, Sekretenko13-fluct, Gavrilov16-helix}).  This
implies certain seed inhomogeneities that cannot be made arbitrarily
small; in other words, the spatial symmetry is broken
\emph{explicitly}.  By contrast, the new mechanism of spin pattern
formation considered here is truly \emph{spontaneous} and takes place
within indefinitely large spatial areas.  In this respect it resembles
the recently reported chimera states in lasers with time-delayed
optoelectronic feedback~\cite{Larger15}.

Recently we have found that a two-dimensional (2D) polariton system
can exhibit spatiotemporal chaos~\cite{Gavrilov16}.  In this work we
find out that a quasi-one-dimensional (1D) microcavity wire arranges
itself into a network of spin-up and spin-down domains alternating
each other in a strict order.  Furthermore, if a particular spin in
such a chain is reversed manually, e.\,g., by means of an additional
properly focused laser beam, under certain conditions all other spins
also get reversed with time, no matter how remote they are.  Thus, a
confined quasi-1D polariton system behaves rather like a stiff lattice
than a fluid: the entire spin network can be reversed by switching one
of its individual nodes.

Paradoxically, turbulence (chaoticity) goes hand in hand with strong
spatial ordering.  To clarify this point, notice that under resonant
plane-wave driving the polariton condensate is usually phase-locked
with respect to the external field, in analogy to a simple damped
pendulum.  All small fluctuations in the vicinity of a given steady
state decay exponentially, whereas sufficiently strong fluctuations
may only trigger a switch into another plane-wave
state~\cite{Liew08-prl-neur, Johne09, Gavrilov12-prb-strains,
  Gavrilov14-jetpl-en}.  Such externally imposed ordering of the
\emph{multistable} polariton system (with sharp switches in singular
points) was long thought to be the sole possibility. It turns out,
however, that the plane-wave states may lose stability and thus become
unfeasible all together in a finite range of pump powers.  The
condensate is then \emph{forbidden} to match the symmetry of the
external field.  As a result, the system gets rid of strict phase
locking and the possibilities open up for both ceaseless variation in
a constant environment (\cite{Gavrilov16}) and the
secondary---internal---ordering of the system.  The spin networks
considered here represent an instance of this novel class of
coherently excited yet internally ordered Bose condensates which
emerge as chimera states even in perfectly homogeneous media.

\textit{Model.}---Right and left circular polarizations of light
correspond to spin-up ($J_z = +1$) and spin-down ($J_z = -1$)
polaritons.  The Gross-Pitaevskii equation
reads~\cite{Kavokin-book-07},
\begin{equation}
  \label{eq:gp}
  i \hbar \frac{\partial \psi_\pm}{\partial t} =
  \left[ \hat E - i \gamma
    + V \psi_\pm^* \psi_\pm^{\vphantom *}
  \right] \psi_\pm^{\vphantom *}
  + \frac{g}{2} \psi_\mp^{\vphantom *}
  + f_\pm^{\vphantom *} e^{-i \frac{E_p}{\hbar} t},
\end{equation}
where the pump and cavity-field amplitudes, $f_\pm$ and $\psi_\pm$,
are spinor functions of time $t$ and spatial coordinates $x, y$ in the
cavity plane.  $V$ is the matrix element of the interaction between
parallel-spin polaritons in the dilute-gas
approximation~\cite{Ciuti98,Vladimirova10,Sekretenko13-10ps}.  Setting
$V = 1$ determines the units of $\psi$ and $f$.  Next, $\gamma$ is the
decay rate; $g$ is the spin coupling rate.  For simplicity, let the
in-plane dispersion law be purely parabolic,
$\hat E = E_0 - \hbar^2 \nabla^2 / 2m$, which is justified near the
low-polariton branch bottom~\cite{Yamamoto-book-2000}.  The pump wave
has frequency $E_p / \hbar$ and zero in-plane wave number ($k = 0)$.

\textit{Solutions beyond multistability.}---When the pump amplitude is
constant in space and time, it is natural to seek the solutions of
Eq.~(\ref{eq:gp}) in the one-mode form
$\psi_\pm(t) = \bar \psi_\pm e^{-i E_p t / \hbar}$.  This leads to
coupled cubic equations for steady-state amplitudes $\bar \psi_+$ and
$\bar \psi_-$.  The solution can be many-valued function of $f_\pm$,
which is referred to as bi- or multistability~\cite{Baas04-pra,
  Gippius07, Gavrilov10-en, Paraiso10}.  Let $f_+ = f_- = f$, so that
the equations for $\psi_+$ and $\psi_-$ become merely the same. It is
well known and experimentally verified that the strict spin symmetry
of this system can break down spontaneously at
$g \gtrsim \gamma$~\cite{Gavrilov13-apl,Gavrilov14-prb-j}.  As a
result, the condensate acquires very high circular polarization (still
being homogeneous in space).  For instance, it could be easily seen
that the one-mode equations are satisfied at
$\bar \psi_- / \bar \psi_+ \to 0$ when
$V|\bar \psi_+|^2 = E_p - E_0 + g/2$ and $\gamma \to 0$; here and in
what follows we consider the case of positive pump detuning
$D = E_p - E_0$.  One can investigate stability of the one-mode
solutions by calculating the spectrum $\tilde E(k)$ of weak
``above-condensate'' excitations depending on
$\bar \psi_\pm$~\cite{Gavrilov16,Gavrilov17-jetpl-en}.  Since
$|\bar \psi_+| \gg |\bar \psi_-|$ or vice versa, the minor spin
component can be neglected.  Then the standard linearization procedure
introduced by Bogolyubov~\cite{Bogolyubov47} yields the following
result,
\begin{gather}
  \label{eq:lambda}
  \tilde E = E_p -i\gamma \pm \frac12 \sqrt{P \pm \sqrt{Q}},\\
  \intertext{where}
  \label{eq:P}
  P = 2 \delta^2 + 4 \delta \chi^2 + 3 \chi^4 + \frac{g^2}{2},\\
  \label{eq:Q}
  Q = \left(4 \delta \chi^2 + 3 \chi^4 \right)^2
  + g^2 \left( 4 \delta^2 + 8 \delta \chi^2 + 3 \chi^4 \right),\\
  \label{eq:designations}
  \delta = \frac{\hbar^2 k^2}{2m} - D, \quad \chi^2 = V|\bar \psi|^2.
\end{gather}
A one-mode solution is unstable when $\Im \tilde E > 0$ for any
$k$. Two different types of instability exist. The first takes place
when $Q > 0$ but $P \pm \sqrt{Q} < 0$, which represents the direct
two-particle scattering of polaritons from the condensate into pairs
of Bogolyubov modes.  (Notice that processes of this general type are
also responsible for the spin symmetry breaking.)  The instability of
the second type occurs at $Q < 0$.  Here the spin coupling and pair
interaction hybridize; the scattered signal/idler modes can have the
same wave number $k=0$ and always have different energies
$\Re \tilde E$ and polarizations: their filling acts to bring back the
spin component absent in the condensate state.  The instability of the
second type destroys the spin-asymmetric solutions, and eventually no
one-mode solutions at all remain stable.  As a result, the field has
to become ceaselessly varying and/or spatially inhomogeneous; in the
general case it exhibits spatiotemporal chaos. The inequalities
$\Im \tilde E > 0$, $Q < 0$ can be satisfied in a finite interval of
$f$ at $g \gtrsim 4\gamma$ and $g/2 \lesssim D \lesssim 2g$, which
constitutes the necessary condition for all phenomena discussed in
this work.

\begin{figure}
  \centering
  \includegraphics[width=\linewidth]{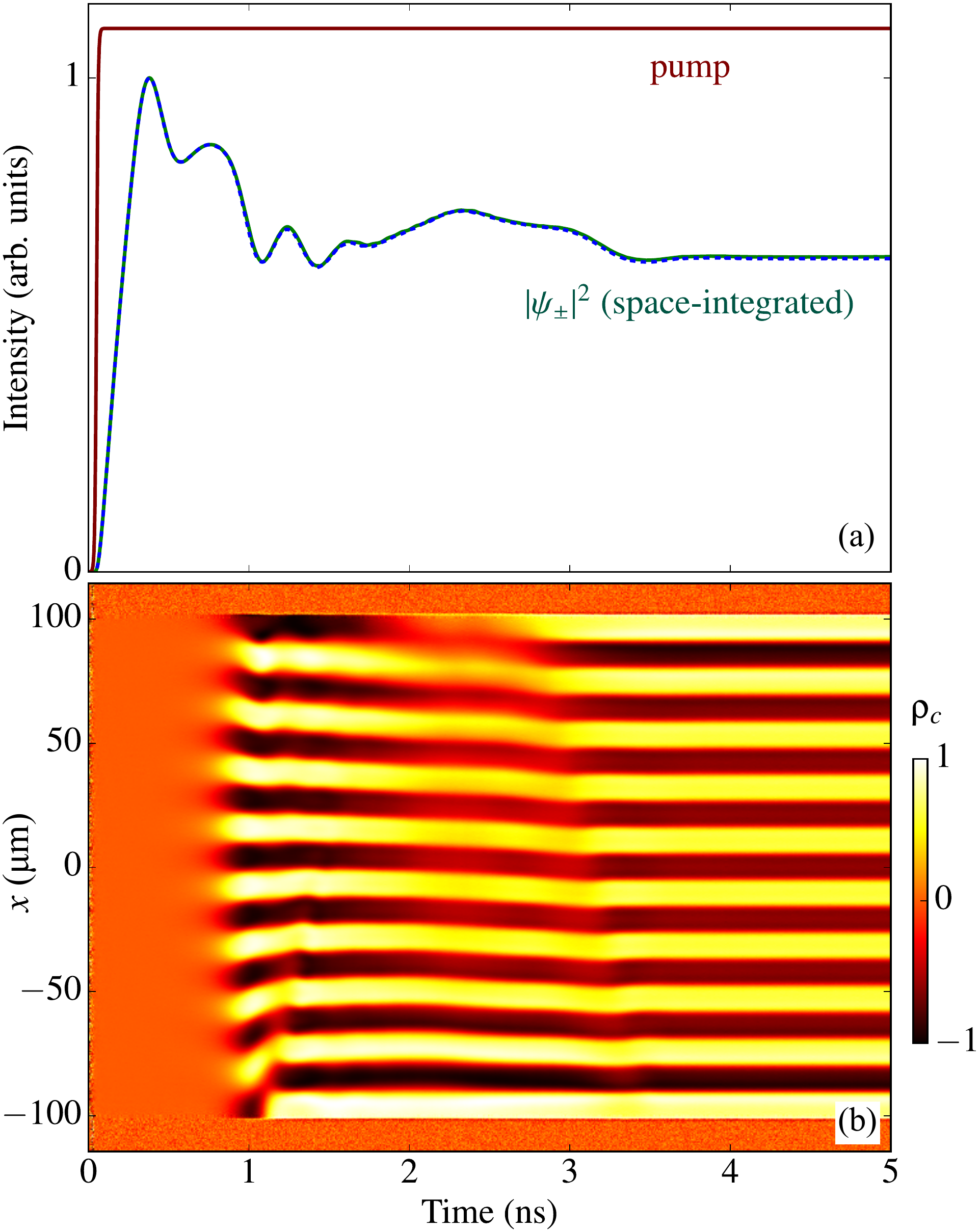}
  \caption{\label{fig:dynamics} Spin pattern formation in a 1D system.
    (a) Time dependences of the pump intensity $|f|^2$ and the
    space-integrated cavity-field components $|\psi_\pm|^2$.  (b)
    Spatiotemporal distribution of the circular-polarization degree
    $\rho_c = (|\psi_+|^2 - |\psi_-|^2) / (|\psi_+|^2 + |\psi_-|^2)$.}
\end{figure}
\begin{figure}
  \centering
  \includegraphics[width=\linewidth]{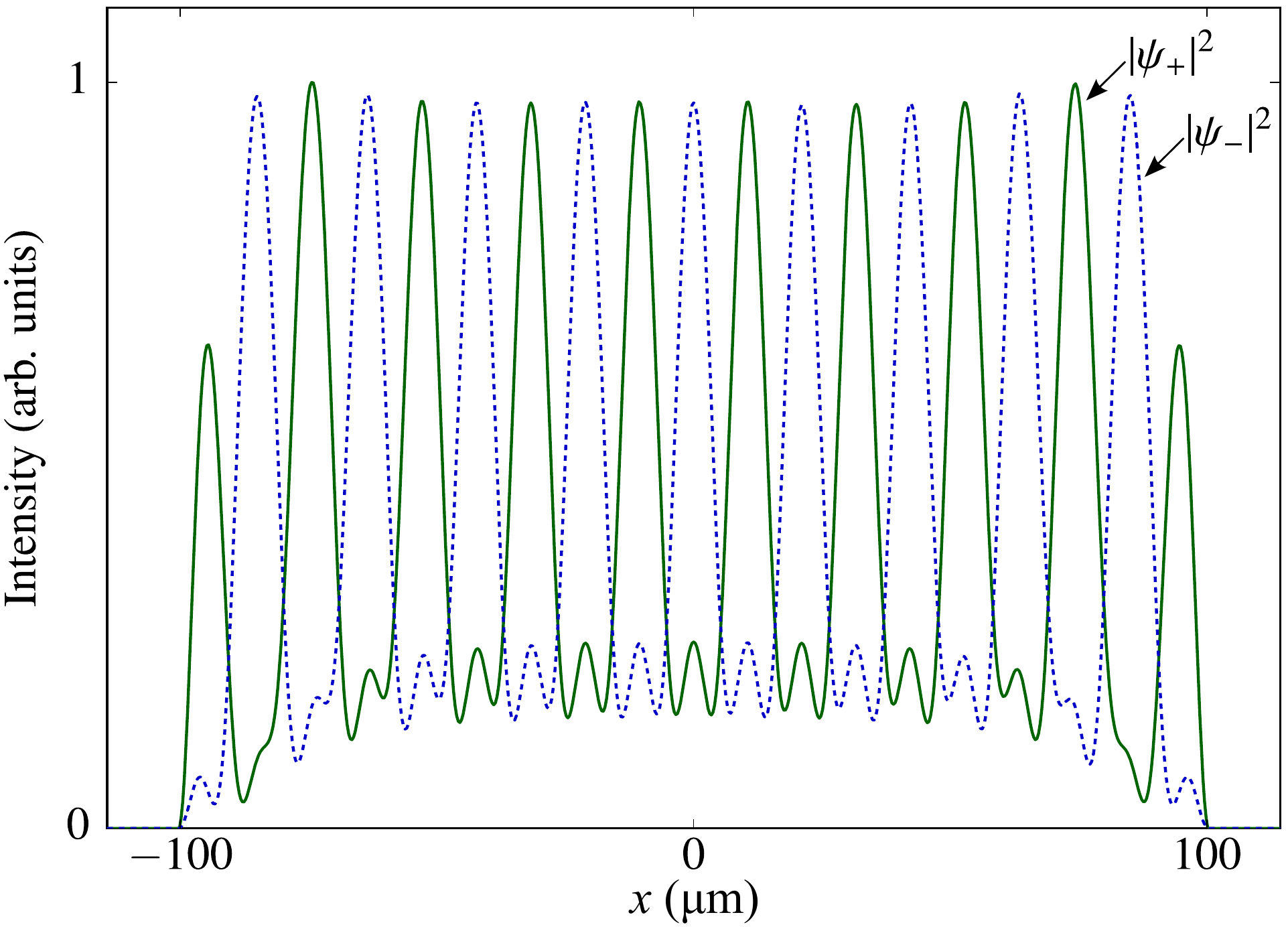}
  \caption{\label{fig:steadystate} Steady-state patterns in a 1D
    system.  Explicit spatial dependences of $|\psi_+|^2$ and
    $|\psi_-|^2$ at the final stage of the evolution displayed in
    Fig.~\ref{fig:dynamics} ($t \gtrsim 4$~ns).}
\end{figure}

\textit{1D wires.}---Let us now turn to a 1D polariton system.  On the
assumption of a zero exciton-photon detuning, the polariton effective
mass $m$ is taken to be twice larger than the photon one:
$m = 2 \epsilon E_0 / c^2$.  The ground-state energy $E_0 = 1.5$~eV
and dielectric constant $\epsilon = 12.5$ are characteristic of
GaAs-based microcavities~\cite{Yamamoto-book-2000}.  The free
parameters are $\gamma = 5~\mu$eV, $g = 50~\mu$eV, and $D = 35~\mu$eV;
they are reachable in state-of-the-art samples and meet the necessary
condition obtained previously.  The length $L$ of the wire amounts to
$200~\mu$m.  On its boundaries, the decay rate $\gamma$ is set to
increase sharply, so that $\psi_\pm$ tend to zero.  The considered
phenomena are qualitatively independent of $L$, provided it is large
enough.

Figure~\ref{fig:dynamics} represents the obtained solution.  The
integral values of $|\psi_+|^2$ and $|\psi_-|^2$ evolve synchronously
[Fig.~\ref{fig:dynamics}(a)].  However, the spin-up and spin-down
fractions of the field get separated in space in nearly 0.5~ns after
the pump has been switched on.  A comparatively slow self-organization
process, which takes the following 2~ns, results in a periodic spin
distribution~[Fig.~\ref{fig:dynamics}(b)].
Figure~\ref{fig:steadystate} shows the finally established spatial
dependences of $|\psi_+|^2$ and $|\psi_-|^2$.  They are not mutually
equivalent, which is an artifact of finite $L$, however, they have the
same integral intensities.  The sites with high degrees of circular
polarization have comparatively high intensities and are separated
from each other by weakly populated zones.

The size $a$ of the spin domains is connected with their
momentum-space width that, in turn, is limited in accordance with the
energy and momentum conservation laws.  On the assumption that the
two-particle breakup of the driven mode $(0, 0) \to (k, -k)$ is the
only scattering process, we have
$\hbar_{\vphantom m}^2 k_\mathrm{max}^2 / 2m = D + g/2$.  Then the
following rough estimate is derived:
$a_\mathrm{min} \approx 2 / k_\mathrm{max} = 2 \hbar / \sqrt{2 m (D +
  g/2)} \approx 6~\mu$m, which turns out to be only moderately smaller
than the actual size of the domains seen in Fig.~\ref{fig:steadystate}
(${\sim} \, 10~\mu$m).

Notice that in the ``spinless'' system continuously driven at $k = 0$
all steady states must be
homogeneous~\cite{Gavrilov14-prb-b,Gavrilov15}.  In our system,
homogeneous solutions are forbidden.  Stability can be reached only
when all inhomogeneities are balanced, which implies a periodic
spatial distribution of the field.  Then all spin states separated by
the lattice period ($2a$) have the same intensity and phase and are
thereby \emph{synchronized} at each given time moment.  However, in a
different parameter area some or many of them fall out of
synchronization even at $t \to \infty$, so that the entire system
never comes to stability.

\textit{Why the spin chains are chimera states?}---The spontaneous
breaking of spatial symmetry is a well-known phenomenon.  Usually it
is understood in view of extremal principles, when, for instance,
pattern formation minimizes the free energy of the system.  After the
system has reached the global minimum, its collective states are
asymptotically stable and described by order parameters~\cite{Haken75}.
In this respect dynamical chimeras are essentially more complex.  In
terms of oscillator networks, they contain both synchronized
(coherent) and desynchronized (incoherent) parts~\cite{Panaggio15}.
Only in the limiting cases chimeras may collapse into fully ordered
states or become fully turbulent; such transitions have recently been
observed in lasers~\cite{Larger15}.  It is difficult to define a
quantity that could serve as a measure of stability of chimeras in the
general case.  The persistence of the irregular part makes the usual
definition of stability inapplicable.  On the other hand, chimera
states are shown to be statistically robust against random structural
perturbations even when the regular part is nearly
absent~\cite{Yao13}.

\begin{figure}
  \centering
  \includegraphics[width=\linewidth]{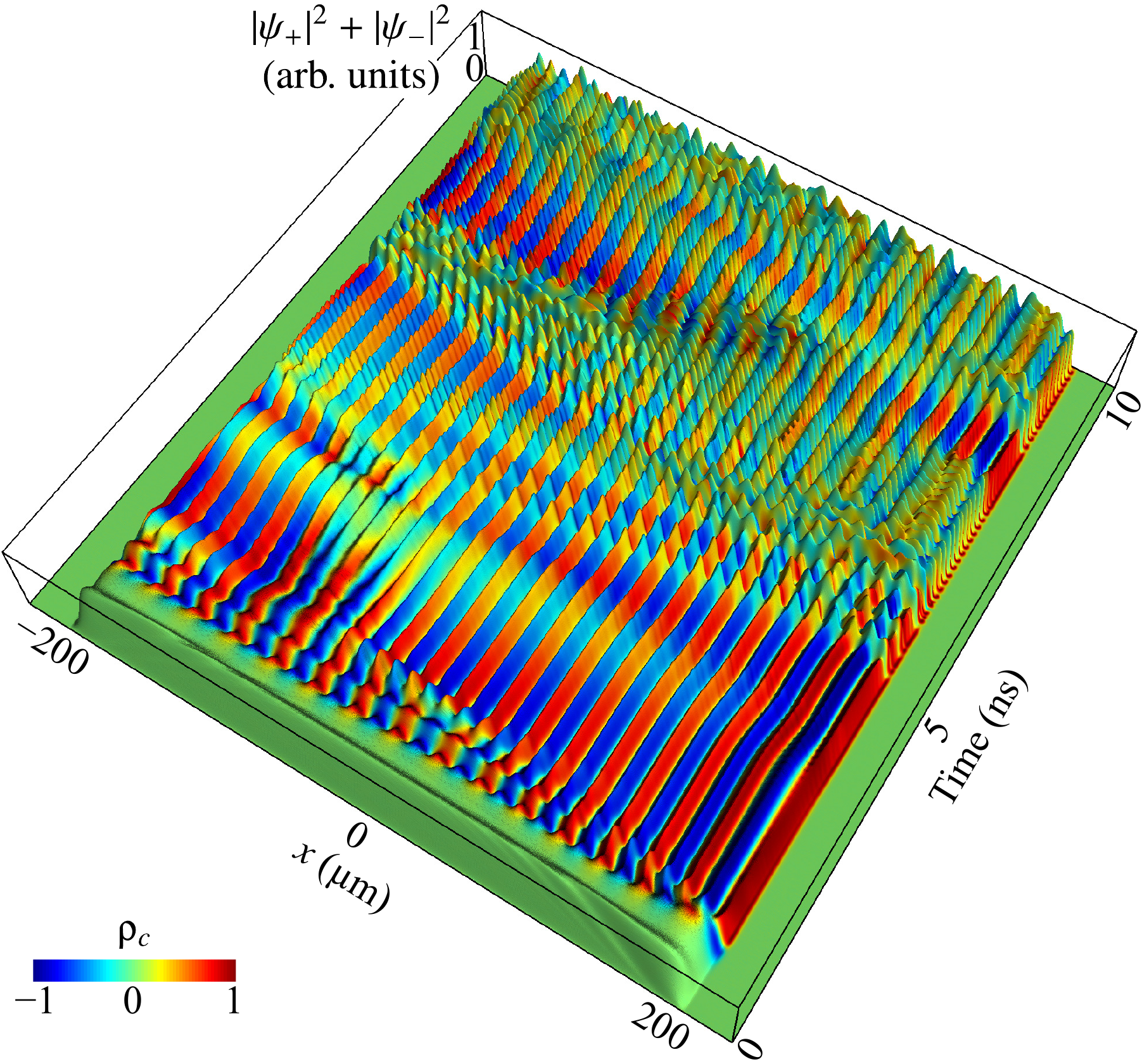}
  \caption{\label{fig:destruction} Formation of a nonstatic chimera in
    a 1D polariton wire.}
\end{figure}

Formation of a nonstatic chimera is shown in
Fig.~\ref{fig:destruction}.  This example is somewhat untypical in
that it combines several dynamical regimes which in their pure states
are observed in separate parameter areas.  Compared to
Fig.~\ref{fig:dynamics}, the calculation is performed for a spatially
longer wire with $L = 400~\mu$m.  The pump $f^2$ is nearly twice
stronger; it is turned on in several tens of picoseconds and then held
constant.

At the first stage the field arranges itself into a set of
opposite-spin domains. Soon after that it behaves more regularly but
exhibits occasional jumps at certain spatial locations.  The
perturbations propagate in space and usually decay with time; the same
effect is also seen in Fig.~\ref{fig:dynamics}.  On the other hand,
the spatiotemporal defects can also give birth to freely
propagating---solitonic---perturbations of the periodic structure.
(Previously, solitons were shown to emerge in the presence of an
artificial periodic potential induced by surface acoustic
waves~\cite{Cerda13}.)  A typical soliton arrives at $x = +200~\mu$m
by $t = 5$~ns.  Solitons also involve local oscillations within the
spin domains they are traveling through; this brings about
\emph{soliton trains}~\cite{Strecker02}.  Multiplying solitons pave
the way for turbulence, however, the system also shows space and time
intervals of comparatively regular (synchronized) evolution, which is
referred to as \emph{intermittency}, a halfway point before real
chaos~\cite{Bohr98}.

This particular example does not end up with full turbulence; in the
future the system behaves similarly to what is seen in the interval
from 5 to 10~ns.  The chimera state remains partially ordered in spite
of all internal perturbations, yet it never becomes static.  In the
general case, the first-order spatial correlation function
$\mathfrak g(d, t)$, which depends on spatial interval $d$ and time
moment $t$, can be less than 1 even for $d=2a$ and $t \to \infty$.
Various static and nonstatic chimera states and, in particular, their
route from perfect periodicity to turbulence upon varying system
parameters, are systematized in Appendix.

\begin{figure}
  \centering
  \includegraphics[width=0.95\linewidth]{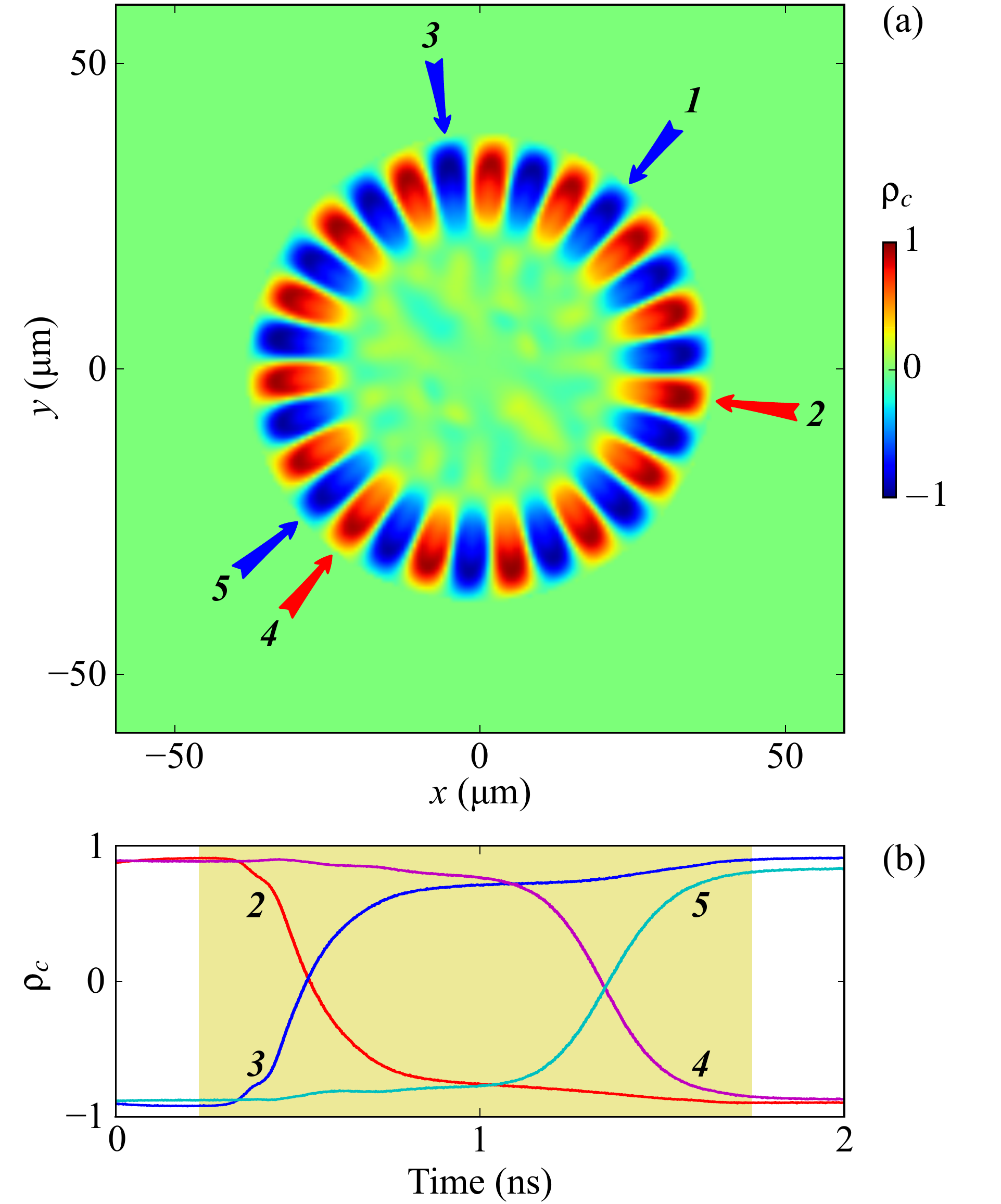}
  \caption{\label{fig:ring} (a) Established spin distribution in a
    homogeneous 2D cavity under ring-shaped excitation.  (b)
    Controlled spin inversion in reference granules~\textit{2--5}.
    The additional laser pulse is focused into a $\mu$m-sized spot at
    position~\textit{1}; it has right circular polarization and acts
    within the shadowed time interval.}
\end{figure}

\textit{2D wires.}---In general, the system does not have to be
strictly one-dimensional to achieve spin granulation.  The periodic
patterns occur equally well in homogeneous 2D cavities, given that
only a narrow (several $\mu$m wide) spatial stripe is pumped from the
outside.  A characteristic example is shown in Fig.~\ref{fig:ring}(a).
The pump has a ring shape, specifically,
$f_+(r) = f_-(r) \propto e^{-(r - R)^2 / 2w^2}$, where $R = 30~\mu$m
and $w = 5~\mu$m.  The system parameters are $\gamma = 20~\mu$eV,
$g = 200~\mu$eV, $D = 150~\mu$eV.  As expected, the established
solution breaks rotational invariance of the model.

\textit{Strong long-range order.}---Two aspects of long-range ordering
should be distinguished.  The first is \emph{predictability}: if one
knows which of two spin states is enhanced at a certain location, all
other sites are thereby also determined.  The second aspect is reduced
to the question of whether an external control over one given spin
state can help manipulate the others.  The answer depends on the
character of the interaction between spatially separated spins.  In
our system, the inter-particle repulsion is definitely local, so one
might suppose, on one hand, that the effect of an externally created
irregularity of the periodic structure should decay with increasing
distance.  On the other hand, self-organization means that all
irregularities are subject to the ``enslaving'' (\cite{Haken75})
forces that keep the system ordered and may even reorder it in
response to changing environment.

The above considerations lead one to the idea of the following
numerical experiment.  Let us take the established system represented
by Fig.~\ref{fig:ring}(a) and perturb it with an additional pump beam
focused into a $1~\mu$m spot in such a way that the spin of a
particular granule is reversed.  The intensity of this beam becomes
negligible already in a few microns away from the target granule so
that it cannot affect remote locations directly.  The calculations
show that after a comparatively short-term perturbation the spin
granules are restored in precisely the same states and positions.  If,
by contrast, the pulse is long enough, then all of the spin states get
reversed one after another; and after the local pulse has gone they
remain steady.

In Fig.~\ref{fig:ring}(a), the granule whose spin is to be reversed
manually is labeled \textit{``1''}.  Labels \textit{2--5} mark the
reference sites whose future dynamics (circular-polarization degree
vs.\ time) is explicitly depicted in Fig.~\ref{fig:ring}(b); the time
span of the additional local pulse is shadowed.  It is seen that
comparatively nearby granules \textit{2} and \textit{3} get reversed
in about 0.3~ns, whereas the switches of \textit{4} and \textit{5}
take ${\sim}\,1$~ns longer.  As a result, all spin states are reversed
in due order, which constitutes a basic prototype of information
transmission.

The considered phenomena strongly depend on transverse dimension $w$.
At large $w$ the field is aperiodic and usually takes the shape of
chaotically placed filaments~\cite{Gavrilov16} resembling turbulent
liquids.  Such systems are long-ordered, but they cannot be
manipulated predictably.  On the contrary, decreasing $w$ involves
strong ordering in the form of a stiff but not necessarily static spin
network.

\textit{Conclusion.}---In summary, it is predicted that resonantly
driven systems of locally interacting bosons can form chimera states
which are different from both the Kuramoto networks
(\cite{Kuramoto02,Abrams04,Acebron05}) and lasers with dime-delayed
feedback (\cite{Larger13,Larger15}).  Driven and dissipative Bose
systems are shown to rid themselves of strict phase locking with
respect to the driving field, which can result in strong internal
ordering and bright solitons propagating in spontaneously formed
periodic domain structures.  Unlike quasi-equilibrium Bose
condensates, the ``incoherent'' part of a polariton chimera state has
purely dynamical nature; the system is not coupled to a thermal
reservoir and thus can be manipulated immediately by optical means.

\begin{acknowledgments}
  I wish to thank V.\,D.~Kulakovskii, S.\,G.~Tikhodeev, and
  N.\,A.~Gippius for stimulating discussions.  The work was supported
  by the Russian Science Foundation (Grant No.\ 16-12-10538).
\end{acknowledgments}

\bigskip

\begin{center}
  \large APPENDIX
\end{center}

Here we systematize the chimera states formed in 1D polariton wires.
In particular, transition from static periodic patterns to disordered
states is illustrated.

\begin{figure*}
  \centering
  \includegraphics[width=0.71\linewidth]{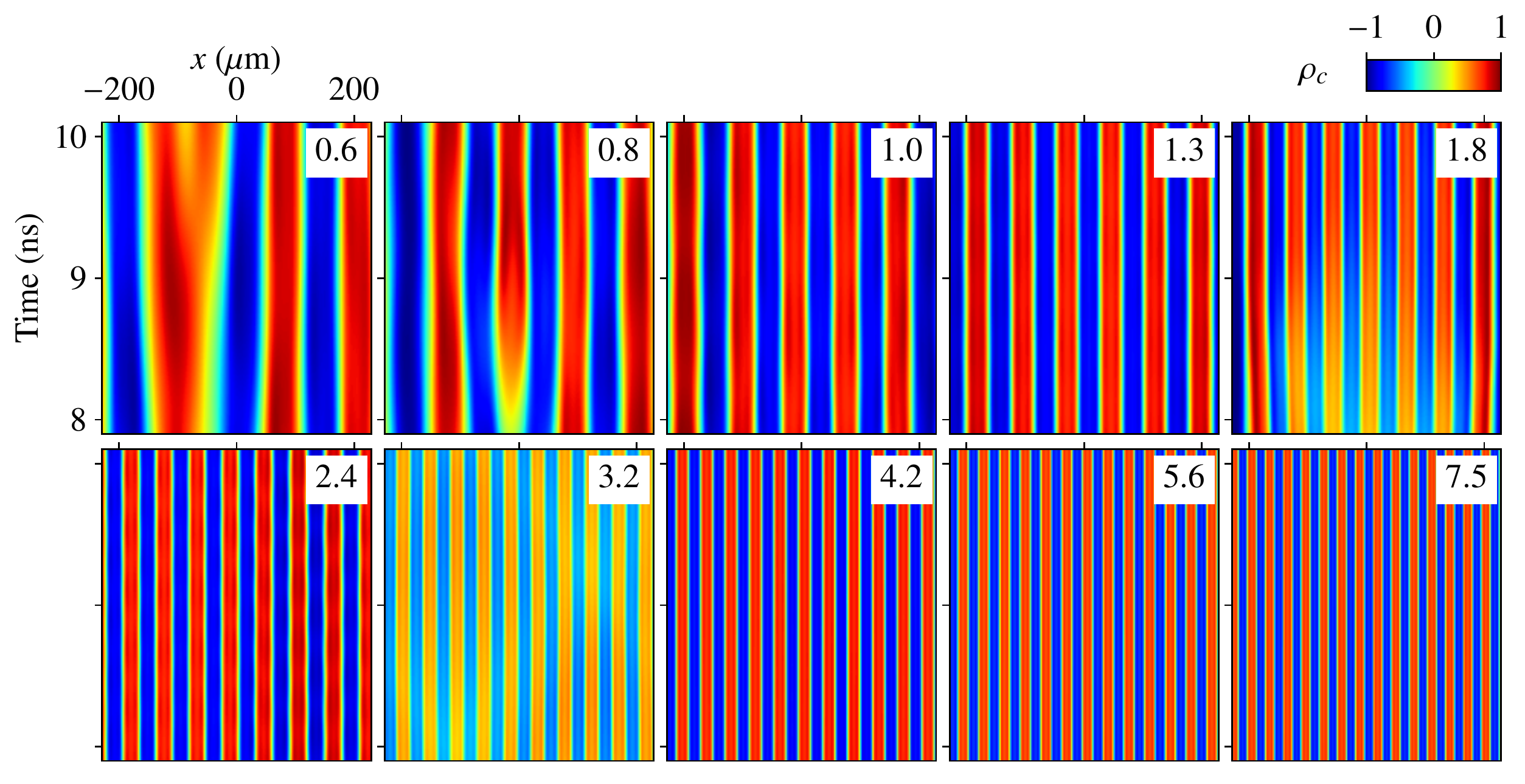}
  \caption{\label{fig:0-0} Chimera states in a polariton wire at
    different $\gamma$ and fixed ratios $g / \gamma = 5$,
    $D / \gamma = 4$.  Numbers indicate $\gamma$ in $\mu$eV; they form
    a geometric progression, $\gamma_n = \frac43 \gamma_{n-1}$. For
    each figure, the pump intensity is set near the instability
    threshold $f_\mathrm{thr}^2 (\gamma, g, D)$.  Color scale
    represents~$\rho_c$.}
\end{figure*}

\begin{figure*}
  \centering
  \includegraphics[width=0.71\linewidth]{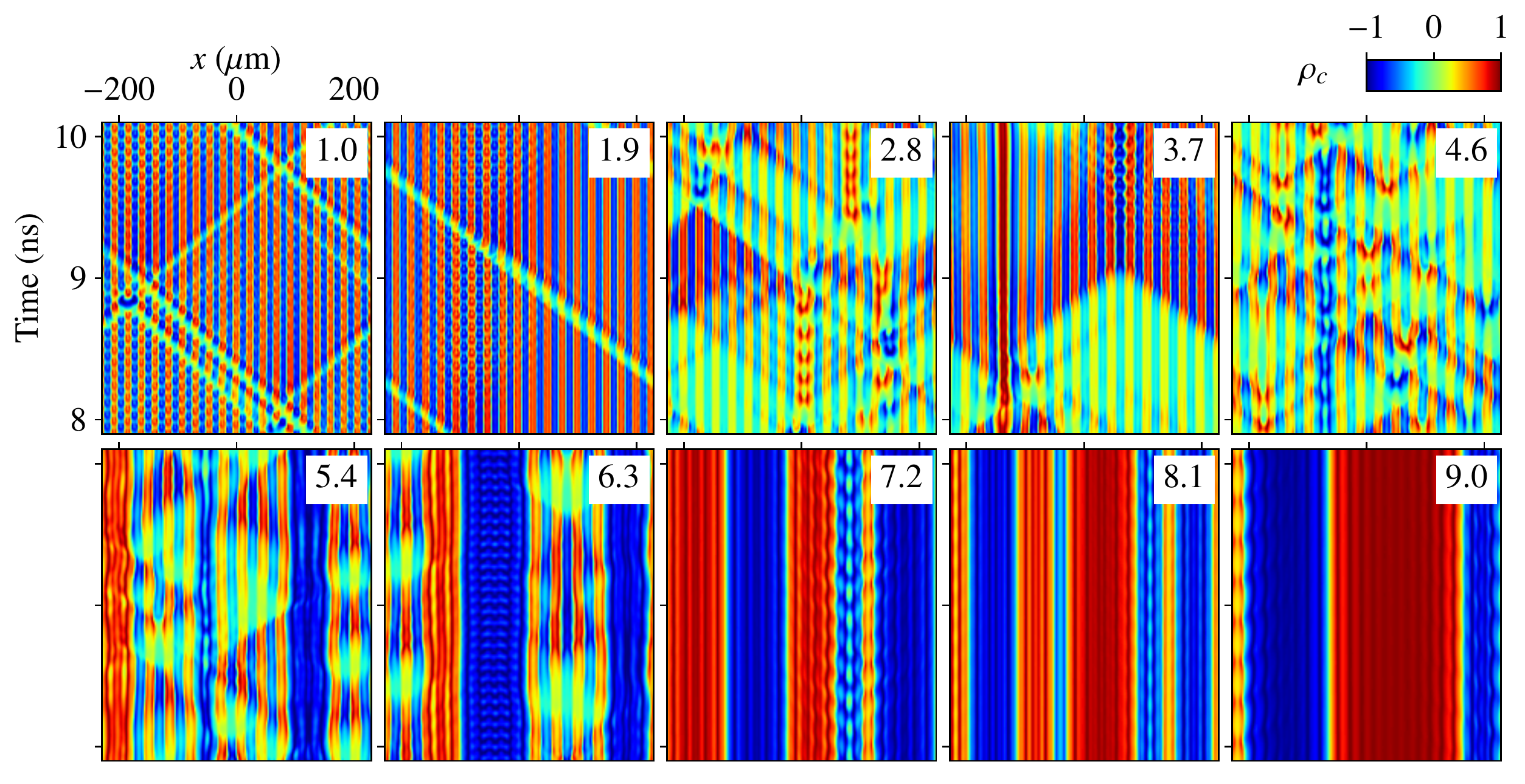}
  \caption{\label{fig:1-0} Chimera states at different pump
    intensities $f^2$ above the instability threshold
    $f_\mathrm{thr}^2$.  Numbers indicate the ratio
    $f^2 / f_\mathrm{thr}^2$. Parameters are $\gamma = 10~\mu$eV,
    $D = g = 5\gamma$.  Color scale represents the degree of circular
    polarization~$\rho_c$.}
\end{figure*}

\begin{figure*}
  \centering
  \includegraphics[width=0.71\linewidth]{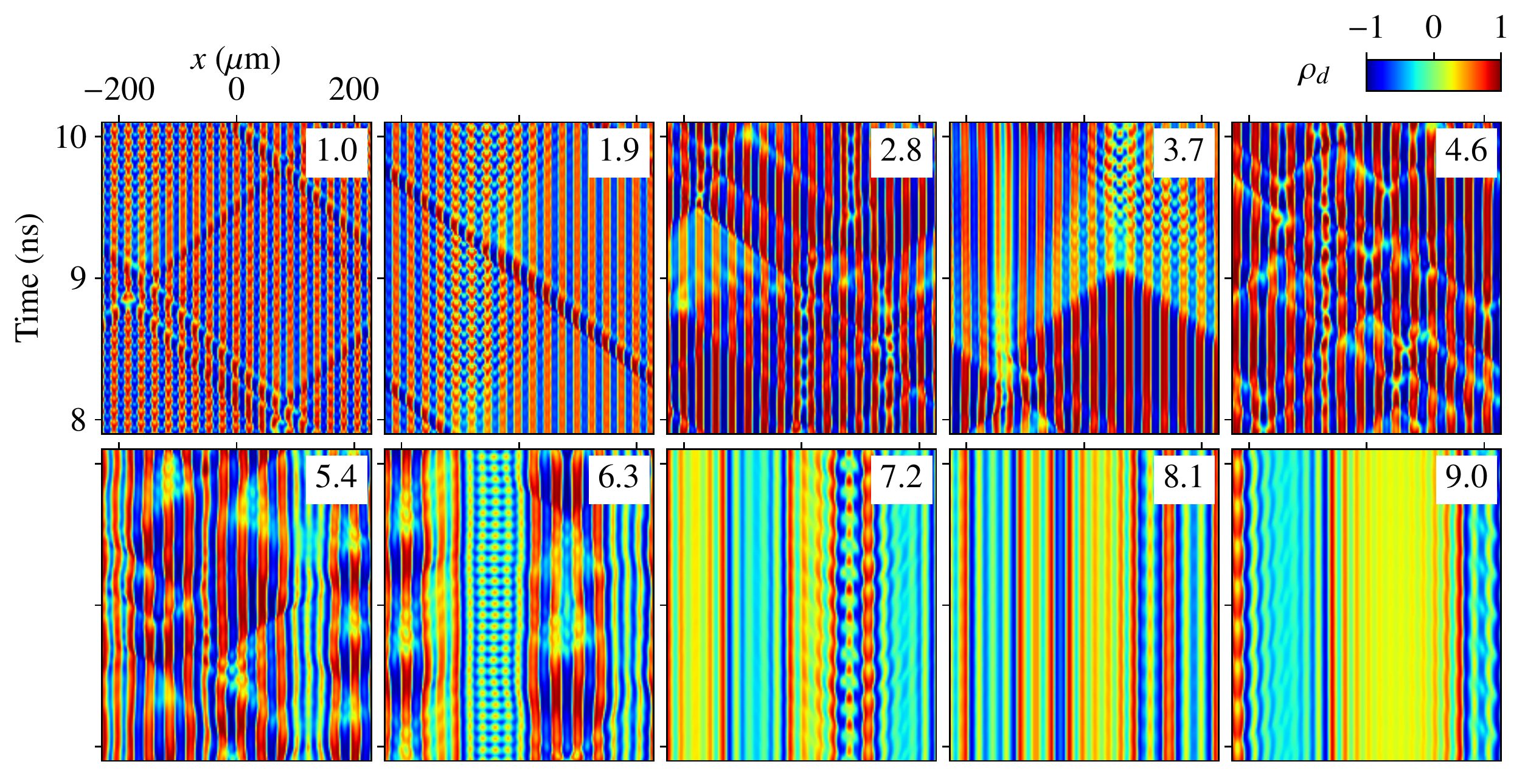}
  \caption{\label{fig:1-1} The same series as in Fig.~\ref{fig:1-0},
    except that color scale represents the degree of the
    $\pm 45^\circ$ linear polarization~$\rho_d$ [see
    Eq.~(\ref{eq:dpd})].}
\end{figure*}

\begin{figure*}
  \centering
  \includegraphics[width=0.71\linewidth]{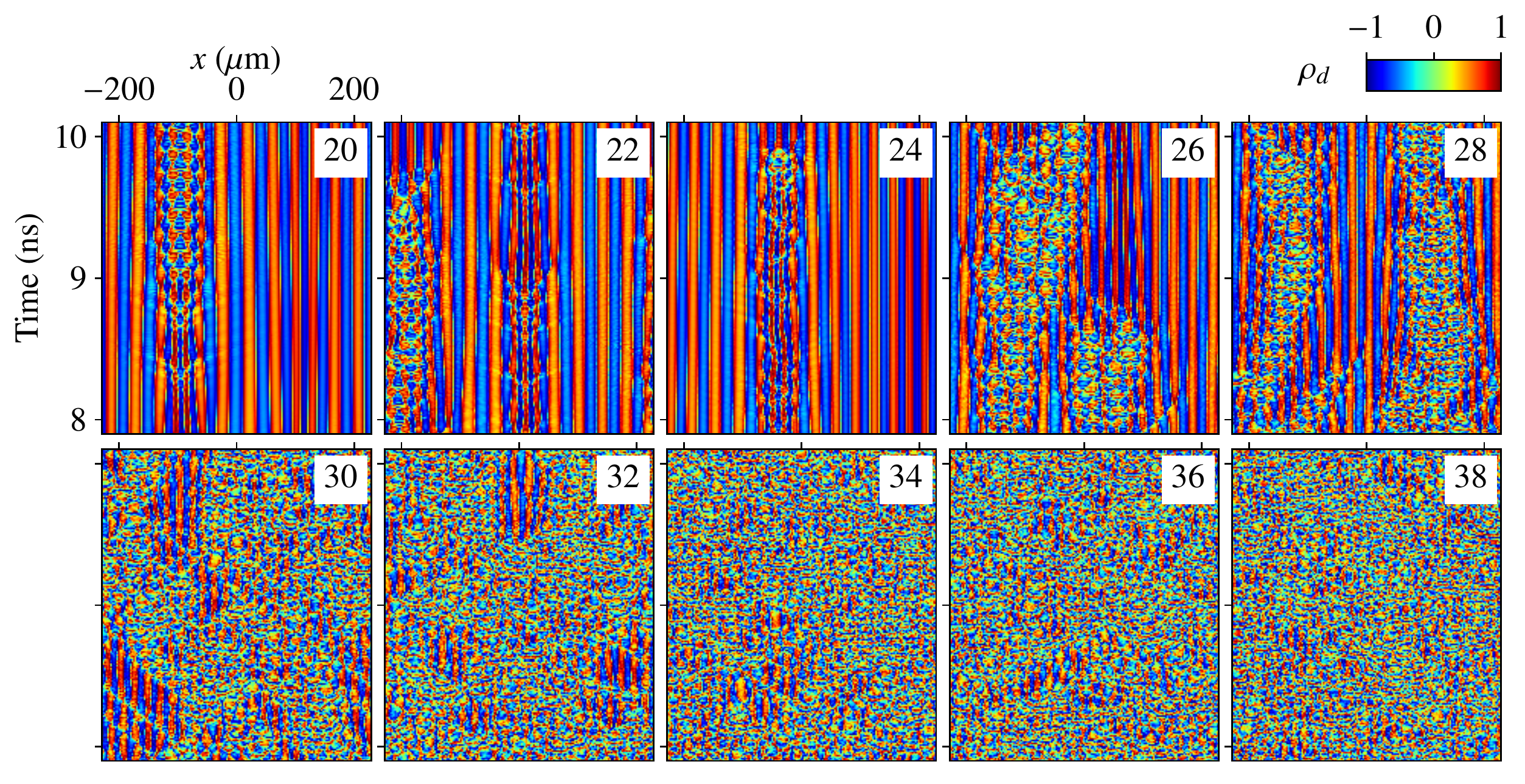}
  \caption{\label{fig:2-1} Chimera states at different $D = g$ and
    constant $\gamma = 5~\mu$eV.  Numbers indicate $g / \gamma$.  For
    each figure, the pump intensity is set in the middle of the
    instability interval.  Color scale represents ~$\rho_d$.}
\end{figure*}

\begin{figure*}
  \centering
  \includegraphics[width=0.70\linewidth]{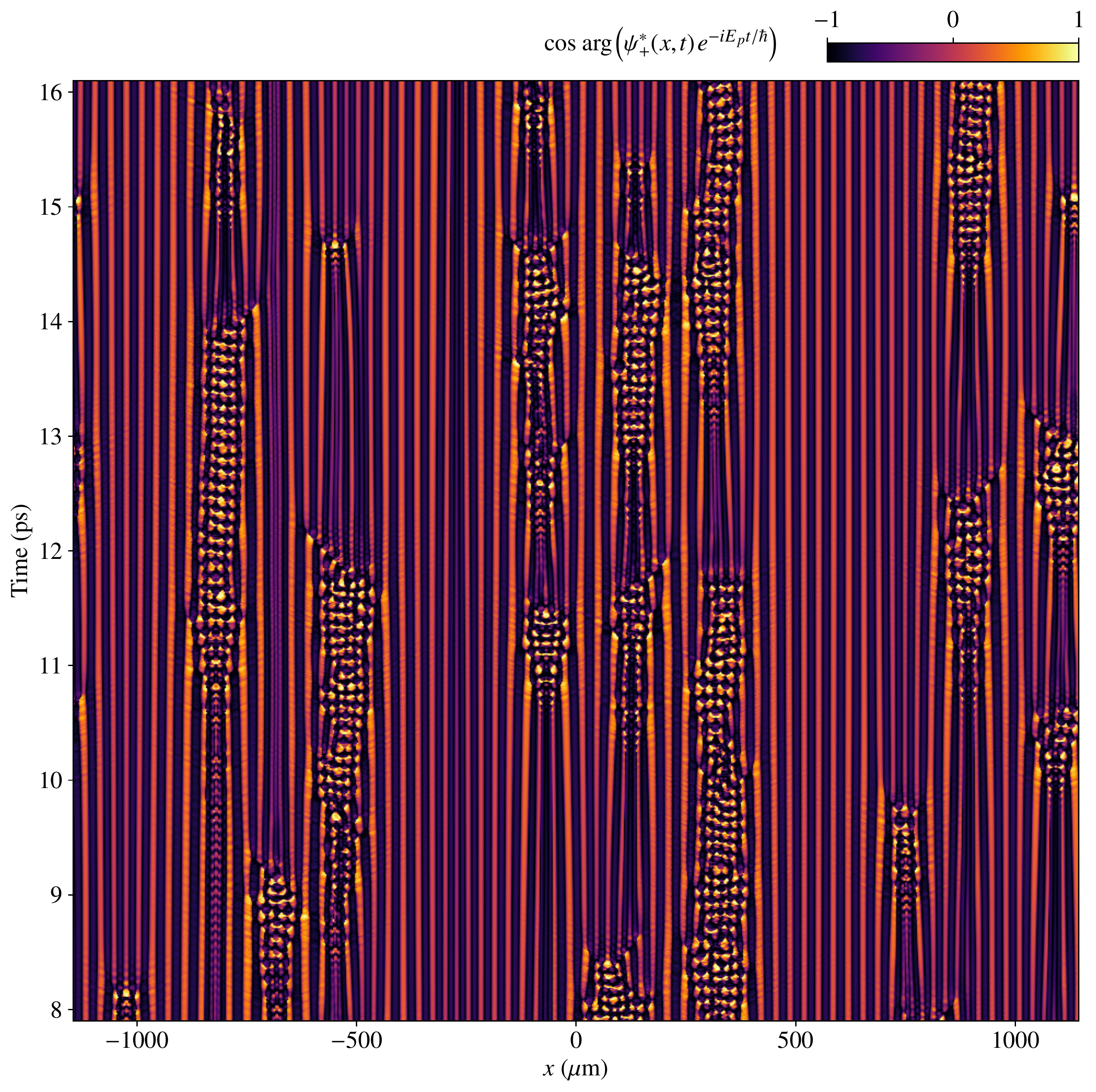}
  \caption{\label{fig:2-wide2} Cosine of the phase difference between
    $\psi_+$ and the driving field. All system parameters are exactly
    the same as in Fig.~\ref{fig:2-1} ($g / \gamma = 24$) except that
    the wire is now 5 times longer, $L \approx 2.3$~mm.}
\end{figure*}

Chimeras appear when the energy splitting $g = E_x - E_y$ of the
polariton eigenstates exceeds their linewidths $\gamma$ by a factor of
4 or greater.  The pump frequency $E_p / \hbar$ should be chosen in
such a way that detuning $D = E_p - E_0 = E_p - (E_x + E_y) / 2$ is
comparable to $g$, specifically, $g/2 \lesssim D \lesssim 2g$.  The
pump polarization should match the upper sublevel, so that
$f_+ = f_-$, $f_x^2 = 2f_+^2$, and $f_y = 0$; this requirement is not
very stiff though.  Then a finite interval of $f^2 \equiv f_x^2$
exists in which no homogeneous solutions of the form
$\psi_\pm(x, t) = \bar \psi_\pm e^{-i E_p t / \hbar}$ remain stable,
be they spin-symmetric ($\bar\psi_+ = \bar\psi_-$) or highly
asymmetric ($|\bar\psi_\pm| \ll |\bar\psi_\mp|$).  In general, this is
valid only up to $g / \gamma \sim 50$; a further increase of
$g / \gamma$ would lead to a new kind of plane-wave multistability
which is not discussed here.

To exclude any transitional effects as much as possible, here and in
what follows we consider the evolution interval starting 8~ns after
the constant pump has been turned on, which exceeds all characteristic
times of the discussed system (e.\,g., $\hbar / \gamma$).  The
boundary conditions are periodic, which allows one to exclude edge
effects.

First, let us demonstrate how to control the size of the spin granules
and, thus, the network period.  In the main part of the article we
have argued that the minimum size $a_\mathrm{min}$ should be sensitive
to the effective mass and pump energy detuning, namely,
$a_\mathrm{min} \approx 2 \hbar / \sqrt{2m (D + g/2)}$.  In the series
displayed in Fig.~\ref{fig:0-0}, parameters $D$, $g$, and $\gamma$ are
successively increased, while the ratios $D / \gamma = 4$ and
$g / \gamma = 5$ are held constant.  The chosen values of $\gamma$ (in
$\mu$eV) are indicated in each subplot, they form a geometric
progression $\gamma_n = \frac43 \, \gamma_{n-1}$.  The color scheme
represents the circular-polarization degree,
\begin{equation}
  \label{eq:cpd}
  \rho_c = \frac
  {\psi_+^*\psi_+^{\vphantom *} - \psi_-^*\psi_-^{\vphantom *}}
  {\psi_+^*\psi_+^{\vphantom *} + \psi_-^*\psi_-^{\vphantom *}},
\end{equation}
as a function of time (within a 2~ns interval) and spatial coordinate.
The pump was set near the threshold
$f^2 \gtrapprox f_\mathrm{thr}^2 (\gamma, g, D)$ for each subplot,
which results in nearly static (collapsed) chimera states. As
expected, the network period $a$ successively decreases.

To obtain nonstatic chimera states, one should to increase field
density.  Figure~\ref{fig:1-0} shows the solutions obtained at
different pump densities $f^2$ above the instability threshold
$f_\mathrm{thr}^2$.  With increasing $f$ the dynamics becomes less
regular, and bright solitons that usually propagate at constant
velocities in perfectly periodic networks become untypical.  Instead,
different spin domains merge and form synchronized clusters.  This
series does not come to turbulence, because at
$f^2 / f_\mathrm{thr}^2 \gtrapprox 9$ the instability interval
terminates and the system comes back to plane-wave multistability;
only one ``cluster'' with a spontaneously chosen but constant
polarization remains afterwards.

The following Fig.~\ref{fig:1-1} shows exactly the same series, but
color now represents the degree of the $\pm 45^\circ$ linear
polarization (sometimes referred to as the third Stokes parameter),
\begin{equation}
  \label{eq:dpd}
  \rho_d = \frac
  {\psi_x^*\psi_y^{\vphantom *} + \psi_y^*\psi_x^{\vphantom *}}
  {\psi_x^*\psi_x^{\vphantom *} + \psi_y^*\psi_y^{\vphantom *}},
\end{equation}
where $\psi_\pm = (\psi_x \mp i \psi_y) / \sqrt{2}$ by definition.
Comparison of Figs.~\ref{fig:1-0} and \ref{fig:1-1} makes clear that
spin-periodic chimera states also show a sort of bistability: for
instance, the spin-up domains have either nearly circular
($\rho_c \sim 1$) or ``diagonal'' linear polarization
($\rho_d \sim 1$) and occasionally switch between these two states.
Solitons propagating through a network with high $|\rho_c|$ have high
$|\rho_d|$ and vice versa.  At the same time, the sign of $\rho_c$
equals the sign of $\rho_d$ at each site and usually remains constant.

Let us now discuss the route to turbulence.  As said above, a mere
increase of the pump intensity would only drive the system beyond the
zone of chimeras and thus make it stable again.  Now we fix the decay
rate $\gamma = 5~\mu$eV and increase both $g$ and $D = g$
(Fig.~\ref{fig:2-1}).  For each $g$, the pump density is set in the
middle of the instability interval.  It is seen that the field
eventually comes to a strongly disordered state. The intervals of a
regular evolution become occasional insertions in a turbulent phase.
The spatial extent and duration of such intervals gradually decrease,
and eventually they get dissolved completely.  This is an instance of
the \emph{intermittent transition to turbulence} whose low-dimensional
prototype was found in the Lorenz system [Commun.\ Math.\ Phys.\
\textbf{74}, 189~(1980)].

The intermittent solutions turn out to be analogous to \emph{discrete}
oscillator networks.  Indeed, a wire of length~$L$ exhibits
$N = L / a$ peaks of the field density.  In a steady state the chain
is periodic: all co-polarized peaks share the same phase and, thus,
are perfectly synchronized.  Increasing density makes them fluctuate,
and their synchronization becomes imperfect (yet still strong).
However, with increasing $g / \gamma$, a number of sites
\emph{completely} fall out of synchronization for quite a long time
but occasionally come back, which is seen in Fig.~\ref{fig:2-wide2}
representing the phase dynamics explicitly.  The desynchronized
domains look like impurities in a periodic lattice, and they hamper
signal transmission.  Their average number per unit length is nearly
constant in time and independent of $N$ (at $N \to \infty$), which is
an important feature of the Kuramoto networks.  Surprisingly, here a
similar network is shown to arise out of a \emph{homogeneous} system
of \emph{locally} interacting bosons.

%


\end{document}